\documentstyle[epsfig,cite]{mn2e}
\begin{document}

\title[{\sl {\sl XMM-Newton}} spectroscopy of high redshift quasars]
    {XMM-Newton spectroscopy of high redshift quasars}
\author[Brocksopp et al.]
    {C.~Brocksopp$^1$\thanks{email: cb4@mssl.ucl.ac.uk}, E.M.~Puchnarewicz$^1$, K.O.~Mason$^1$, F.A.~C\'ordova$^2$, W.C.~Priedhorsky$^3$\\
$^1$Mullard Space Science Laboratory, Holmbury St. Mary, Dorking, Surrey RH5 6NT\\
$^2$Department of Physics, University of California, Riverside, CA, 92521\\
$^3$Los Alamos National Laboratory, Los Alamos, NM 87545\\
}
\date{Accepted ??. Received ??}
\pagerange{\pageref{firstpage}--\pageref{lastpage}}
\pubyear{??}
\maketitle

\begin{abstract}
We present {\sl XMM-Newton} X-ray spectra and optical photometry of four high redshift ($z=2.96--3.77$) quasars, [HB89]~0438$-$436, [HB89]~2000$-$330, [SP89]~1107+487 and RX~J122135.6+280613; of these four objects the former two are radio-loud, the latter two radio-quiet. Model fits require only a power law with Galactic absorption in each case; additional intrinsic absorption is also needed for [HB89]~0438$-$436 and RX~J122135.6+280613. The spectra are hard ($\Gamma \sim 1.7$ for [HB89]~0438$-$436, [HB89]~2000$-$330 and $\sim 1.4$ for RX~J122135.6+280613) with the exception of [SP89]~1107+487 which is softer ($\Gamma\sim 2.0$); the combined Galactic and intrinsic absorption of lower energy X-rays in the latter source is much less significant than in the other three. The two intrinsically unabsorbed sources have greater optical fluxes relative to the X-ray contributions at the observed energies. While there is no need to include reflection or iron line components in the models, our derived upper limits (99\% confidence) on these parameters are not stringent; the absence of these features, if confirmed, may be explained in terms of the high power law contribution and/or a potentially lower albedo due to the low disc temperature. However, we note that the power-law spectrum can be produced via mechanisms other than the Comptonization of accretion disc emission by a corona; given that all four of these quasars are radio sources at some level we should also consider the possibility that the X-ray emission originates, at least partially, in a jet.
\end{abstract}


\section{Introduction}

The first quasars were discovered as a result of radio surveys in the 1950s, with searches extending out to $z \approx 5$ by the mid 1990s (see e.g. Peterson 1997 and references therein) and more recently to $z=6.43$ (Fan et al. 2003). However it was not until much more recently, with the advent of the e.g. {\sl ASCA} satellite, that it became possible to obtain X-ray spectra of these high redshift objects. Such observations are important for probing the central regions of the quasar and determine the mechanism behind their large luminosity output. {\sl XMM-Newton} has now given us the ability to study the spectra of high redshift quasars to an unprecedented level, the first of which was PKS 0537$-$286 at $z=3.10$ (Reeves et al. 2001). The most distant quasar ($z=5.74$) to have been observed with {\sl XMM-Newton} to date is SDSSp J104433.04-012502.2 (Brandt et al. 2001).

Despite the readily accepted fact that active galaxies and quasars are all powered by accretion on to a black hole there are still many properties of quasars that are not well-understood. The most widely accepted spectral model (for X-ray spectral features observed in low redshift Seyfert galaxies) invokes fluorescence and reprocessing in an accretion disc and predicts a Comptonized reflection bump peaking at $\sim30$ keV, as well as an iron K$\alpha$ emission line at $\sim$ 6.4--6.97 keV (e.g. Lightman \& White 1988; George \& Fabian 1991; Fabian et al. 2000; Reynolds \& Nowak 2003). However, as we discuss further in Section 5, there are other models that can describe the data equally well and are worthy of further study.

We have obtained {\sl XMM-Newton} X-ray and optical data for four quasars -- [HB89]~0438$-$436, [HB89]~2000$-$330, [SP89]~1107+487 and RX~J122135.6+280613 with redshifts in the range 2.96--3.77 -- with the intention of probing quasar emission to rest energies as high as $\sim 50$ keV and comparing radio-loud and radio-quiet quasars (where we use the definition of radio-loud as given in Bechtold et al. 1994a). In the remainder of this section we summarise previous observations of these sources. We then describe and discuss our observations and resultant lightcurves and spectra in the remainder of the paper.

\begin{table*}
\caption{Table listing the four quasars, their positions, Galactic column density (from Dickey \& Lockman 1990), redshifts, date of observations and exposure times (time on-source) of the X-ray observations with {\sl XMM-Newton}.}
\begin{tabular}{llcccclc}
\hline
\hline
Quasar & Alternative Name&Right Ascension & Declination & Galactic N$_{\mbox{\sc h}}$ & Redshift & Obs. Date & Time On-Source\\
&&(J2000)&(J2000)&($10^{20}$cm$^{-2}$)&&&(ks)\\
\hline
\mbox{[HB89]} 0438$-$436&PKS 0438$-$43&04 40 17.2   &$-$43 33 08.6  &1.75&2.852&2002 April 6 &9.9\\
\mbox{[HB89]} 2000$-$330&PKS 2000$-$330&20 03 24.1   &$-$32 51 45.1  &7.89&3.773&2002 April 14&21.5\\
\mbox{[SP89]} 1107+487&QSO B1107+487&11 10 38.5 &48 31 15.5&1.26&2.96&2002 June 1&31 \\
RX J122135.6+280613&RIXOS 126$-$27&12 21 35.6 &28 06 14.1&1.9&3.305&2002 June 26&40.4\\
\hline
\end{tabular}
\label{quasars}
\end{table*}

\begin{figure*}
\begin{center}
\leavevmode
\psfig{file=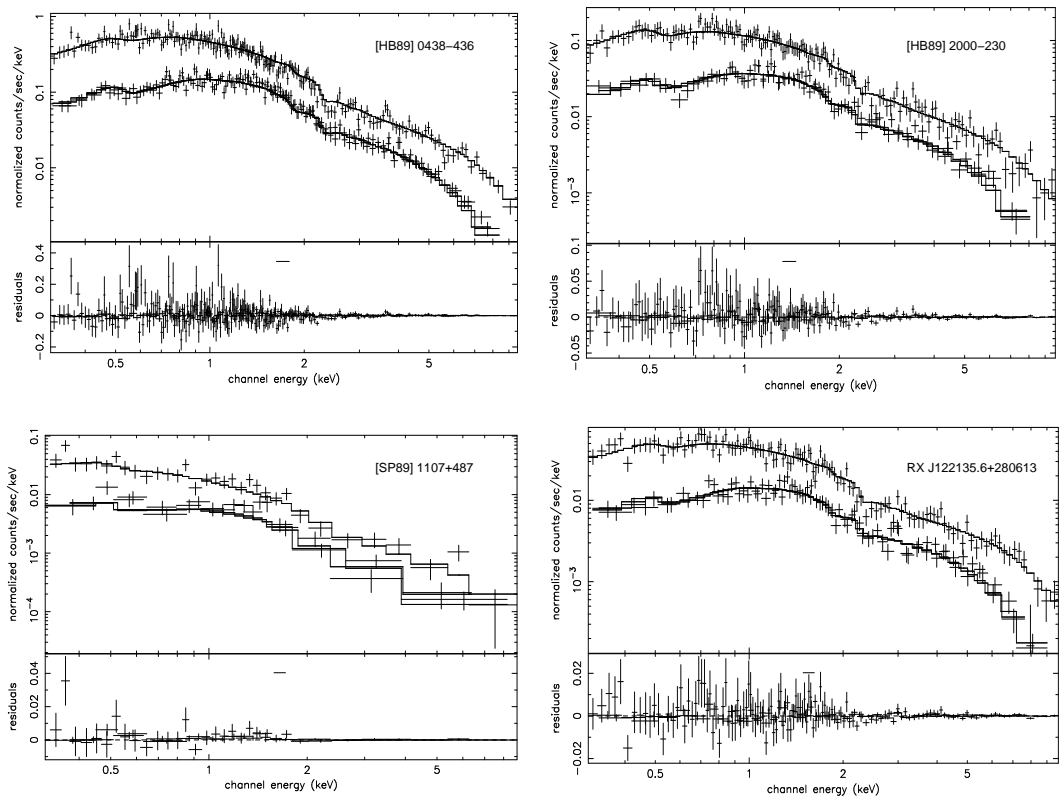,angle=0,width=14cm}
\caption{Combined PN/MOS spectra for each of the four quasars overplotted by the best-fit model, which in each case was an absorbed power-law; an intrinsic absorption component was required in addition to the Galactic value in the case of [HB89] 0438$-$436 and RX J122135.6+280613. The lower panel of each plot shows the residuals of fit in units of $\sigma$ and fit parameters are listed in Table~\ref{model}. Each residuals panel shows a horizontal line which indicates the range searched for iron lines in the observed frame (see text).}
\label{spectra}
\end{center}
\end{figure*}

\subsection{[HB89] 0438$-$436 ($z=2.852$)}

[HB89]~0438$-$436 is a radio-loud (7.6 Jy at 5 GHz; see e.g. Cappi et al. 1997) and relatively well-studied quasar, having been the first high redshift source for which an X-ray spectrum was obtained (Wilkes et al. 1992). Contrary to the majority of quasars observed previously, the spectrum appeared to have a redshifted absorption component, dismissing the suggestion that the degree of intrinsic absorption was inversely proportional to the luminosity. Cappi et al. (1997) analysed the combined {\sl ASCA} and {\sl ROSAT} observations and found that a simple model consisting of a power-law ($\Gamma =1.48\pm0.1$, $n_H=n_{H, Gal}+4.34\times10^{20}$ cm$^{-2}$) and intrinsic cold absorption was most appropriate.

\subsection{[HB89] 2000$-$330 ($z=3.773$)}

[HB89]~2000$-$330 is another radio-loud (1.2 Jy at 5 GHz; see e.g. Hirabayashi et al. 2000) quasar which has been observed with {\sl ROSAT}. This quasar was also modelled with a power-law ($\Gamma$ fixed at $\sim 1.7$) and was found to need no additional absorption component above the (relatively large) Galactic value (Elvis et al. 1994).

\subsection{[SP89] 1107+487 ($z=2.96$)}

[SP89]~1107+487 was identified as a possible quasar from optical spectra during the Case Low-Dispersion Northern Sky Survey (Sanduleak \& Pesch 1989). It is radio-quiet but has been detected at 5 GHz with a flux density of $\sim 0.5$ mJy (Kuhn et al. 2001). The source was observed by {\sl ROSAT} in 1992 and its spectrum obtained ($\Gamma \sim 1.6 - 2$; Bechtold et al. 1994b.

\begin{table*}
\caption{Table of optical and ultraviolet photometry from the Optical Monitor observations.}
\label{optical} 
\begin{tabular}{lcccc}
\hline
\hline
&\multicolumn{4}{c}{Filter/magnitudes ($\lambda_{eff}$)}\\
Quasar&$V$ (5483 \AA)&$B$ (4443 \AA)&$U$ (3735 \AA)&$UVW1$ (2910 \AA)\\
\hline
\mbox{[HB89]} 0438$-$436&$19.91\pm0.29$&$20.74\pm0.23$&$>22.0$& --\\
\mbox{[HB89]} 2000$-$330&$17.59\pm0.12$&$18.91\pm0.14$&--&--\\
\mbox{[SP89]} 1107+487  &$17.01\pm0.02$&$17.31\pm0.01$&$17.61\pm0.02$&$19.78\pm0.65$\\
RX J122135.6+280613     &$19.58\pm0.16$&$20.25\pm0.12$&$21.50\pm0.44$&--\\
\hline
\end{tabular}
\end{table*}

\subsection{RX J122135.6+280613 ($z=3.305$)}

RX~J122135.6+280613 is a radio-quiet quasar which was discovered by {\sl ROSAT} as part of the {\sl ROSAT} International X-ray/Optical Survey (Mason et al. 2000); the optical counterpart was also detected. No X-ray spectra have been obtained previously. The radio source was detected at a flux density of $\sim 29$ mJy at 1.4 GHz in the FIRST survey (e.g. White et al. 1997 and references therein).

\section{{\sl XMM-Newton} Observations}

{\sl XMM-Newton} observed the four targets in 2002 April ([HB89]~0438$-$436, [HB89] 2000$-$330) and June ([SP89] 1107+487, RX J122135.6+280613) and obtained images with each of the three X-ray EPIC cameras (PN, MOS1, MOS2); details of the sources and their observations are listed in Table~\ref{quasars}.  The EPIC full frame mode and thin filter were used. Additional images were obtained using the Optical Monitor camera through the $U$, $B$ and $V$ filters and also through the $UVW1$ filter for [SP89] 1107+487. 

The PN and MOS spectra were reduced using the SAS v5.3.3 version of {\sc epchain} and {\sc emchain} reduction pipelines respectively. There were no significant background flares. The OM data were reduced using the SAS v5.4 {\sc omichain} reduction pipeline. The photometry was measured from the OM images using the aperture photometry routine within {\sc gaia}, typically using an aperture of $\sim 1.5$ times the FWHM, an inner radius for the sky annulus of $\sim 4$ times the FWHM and width of the annulus of $\sim 5$ pixels; we list the resultant photometry in Table~\ref{optical}. Finally the magnitudes were flux calibrated with respect to Vega.

\section{X-ray Lightcurves}
 
A 0.2--10 keV lightcurve was obtained from the PN data of each quasar by using the task {\sc xmmselect} to extract a circular region around the source. Further circular regions were extracted from source-free areas of sky to provide a background sample. All single--quadruple events (i.e. pattern 0--12) with quality flag 0 were included. Finally the background lightcurve was subtracted from the source lightcurve to a resolution of 100 seconds ([HB89]~0438$-$436, [HB89] 2000$-$330), 200 seconds (RX J122135.6+280613) or 500 seconds ([SP89] 1107+487) using {\sc lcmath}. Each lightcurve is featureless and consistent with statistical noise only -- the results of fitting zero-order polynomials to them can be seen in Table~\ref{lightcurve}.

\begin{table*}
\caption{Values of mean and standard deviation of each lightcurve for the energy range 0.2--10 keV in the observed frame. The second column shows the corresponding rest frame for each quasar. The third column shows the $\chi^{2}$ following a constant-line fit to the data and the last column shows the number of degrees of freedom.}

\begin{tabular}{lcccc}
\hline
\hline
Quasar &Energy& Mean ($\sigma$)&$\chi^{2}$& d.o.f.\\
&(keV)&(cts/sec)&&\\
\hline
\mbox{[HB89]} 0438$-$436&0.8--38.5&0.78 (0.10)&116.06&100\\
\mbox{[HB89]} 2000$-$330&1.0--47.7&0.24 (0.06)&196.14&216\\
\mbox{[SP89]} 1107+487  &0.8--39.6&0.03 (0.01)&52.98&63\\
RX J122135.6+280613     &0.9--43.1&0.10 (0.02)&165.75&203\\
\hline
\end{tabular}
\label{lightcurve}
\end{table*}

\begin{table*}
\caption{Source count rates in the 0.3--10 keV energy range (observed frame) for each of the three EPIC cameras and the observed flux for the 0.5--2.0 and 2.0--10.0 keV energy ranges (for an absorbed power-law fit). Errors in the count rates are included in parentheses.}
\label{counts} 
\begin{tabular}{lccccc}
\hline
\hline
Quasar&PN Count Rate&MOS1 Count Rate&MOS2 Count Rate&Observed Flux&Observed Flux\\
      &(0.3--10 keV)    &(0.3--10 keV)    &(0.3--10 keV)    &(0.5--2.0 keV)&(2.0--10.0 keV)\\
      &$10^{-3}$ cts/s  &$10^{-3}$ cts/s  &$10^{-3}$ cts/s &$10^{-14}$ erg/cm$^2$/s &$10^{-14}$ erg/cm$^2$/s\\
\hline
\mbox{[HB89]} 0438$-$436&845.2(9.7)&277.1(4.8)&265.3(4.7)&82.9&168.9\\
\mbox{[HB89]} 2000$-$330&222.3(3.8)&72.6(1.9)&65.7(1.8)  &21.1&45.6\\
\mbox{[SP89]} 1107+487  &35.8(1.6)&11.4(0.9)&10.7(0.8)   &3.1&4.0\\
RX J122135.6+280613     &94.9(1.7)&30.5(0.9)&29.2(0.9)   &8.4&26.4\\
\hline
\end{tabular}
\end{table*}

\begin{table*}
\caption{Table listing the four quasars and the resultant spectral parameters using an absorbed power-law model. The observed frame energy range was 0.3 -- 10 keV in each case and the rest frame energy range for each source is listed. More complicated models resulted in lower-quality  fits. In each case the fit was obtained for the combined PN and MOS data and errors quoted for the power-law index and redshifted N$_{\mbox{\sc h}}$ are based on 90\% confidence intervals for two interesting parameters. The last column gives the luminosity (WMAP cosmology) for rest frame energies of 2--30 keV.}

\begin{tabular}{lccccccc}
\hline
\hline
Quasar&Rest Frame Energy&Power-Law&Redshifted N$_{\mbox{\sc h}}$&Normalization& $\chi^{2}_{\nu}$ (d.o.f.)&Luminosity\\
&Range (keV)&Index&($10^{22}$ cm$^{-2}$)&($10^{-4}$ cm$^{-2}$s$^{-1}$keV$^{-1}$)&&($10^{47}$ erg/s)\\
\hline
\mbox{[HB89]} 0438$-$436&1.2--38.5&1.74$^{+0.04}_{-0.04}$&1.28$^{+0.20}_{-0.19}$&4.45&1.03 (375)&1.59\\
\mbox{[HB89]} 2000$-$330&1.4--47.7&1.73$^{+0.06}_{-0.06}$&$<0.73$&1.19&0.85 (254)&0.75\\
\mbox{[SP89]} 1107+487&  1.2--39.6&1.96$^{+0.13}_{-0.16}$&$<0.52$&0.15&1.10 (54)&0.05\\
RX J122135.6+280613  &   1.3--43.1&1.41$^{+0.06}_{-0.06}$&0.90$^{+0.68}_{-0.10}$&0.41&0.91 (209)&0.28\\

\hline
\end{tabular}
\label{model}
\end{table*}

\section{X-ray Spectra}

Spectra (0.3--10 keV) were extracted using {\sc xmmselect} in a similar manner to that described above, including photons with pattern 0--4 for the PN and 0--12 for the MOS. We note that the source counts dominated in the range 0.2--0.3 keV also but that this region was omitted because the spectrum response is not well-known. {\sc grppha} was used to bin the data using a minimum of 30 counts in each bin and 20 counts per bin in the case of [SP89] 1107+487. The datasets were then loaded into {\sc xspec}, along with the corresponding response matrices (epn\_ff20\_sdY9\_thin.rsp, m1\_thin1v9q19t5r5\_all\_15.rsp, m2\_thin1v9q19t5r5\_all\_15.rsp for the PN, MOS1 and MOS2 respectively). All our sources were on-axis and so our results are independent of the response matrices used. We show in Table~\ref{counts} the source count rates (0.3--10 keV) for each of the three EPIC cameras and the flux in the 0.5--2.0 and 2.0--10 keV energy ranges (for an absorbed power-law model -- see below).

Cold absorbed power-law models were used to fit the X-ray spectra of all four quasars, firstly for each individual instrument and secondly for the combined PN+MOS data. The shape of the residuals suggested that Galactic absorption alone was insufficient to provide good fits to the data in at least two of the objects. An additional redshifted cold absorption component ($n_{H, z}$) improved the fit for [HB89] 0438$-$436 ($\Delta\chi^2=298.3$ for 1 additional degree of freedom and an Ftest statistic $F$=289; $n_{H, z}=(1.28^{+0.20}_{-0.19})\times 10^{22}\mbox{cm}^{-2}$) and RX J122135.6+280613 ($\Delta\chi^2=28.3$ for 1 additional degree of freedom, $F$=31; $n_{H, z}=(0.9^{+0.68}_{-0.10})\times 10^{22}\mbox{cm}^{-2}$); the inclusion of intrinsic absorption made no significant improvement for [HB89] 2000$-$330 ($n_H \le 0.95\times 10^{22}$ atoms/cm$^2$ with 99\% confidence; $\Delta\chi^2=1.7$ for 1 additional degree of freedom, $F$=2) or [SP89] 1107+487 ($n_H \le 1.5 \times 10^{22}$ atoms/cm$^2$ with 99\% confidence; $\Delta\chi^2=0.1$ for 1 additional degree of freedom, $F$=0.1). Model fits to the combined PN+MOS data and their residuals are plotted in Fig.~\ref{spectra}; the fit parameters are listed in Table~\ref{model}. Additional fits were attempted using more complicated models, such as the inclusion of a blackbody component; there was no improvement over the simple absorbed power-law.

\begin{figure*}
\begin{center}
\leavevmode
\psfig{file=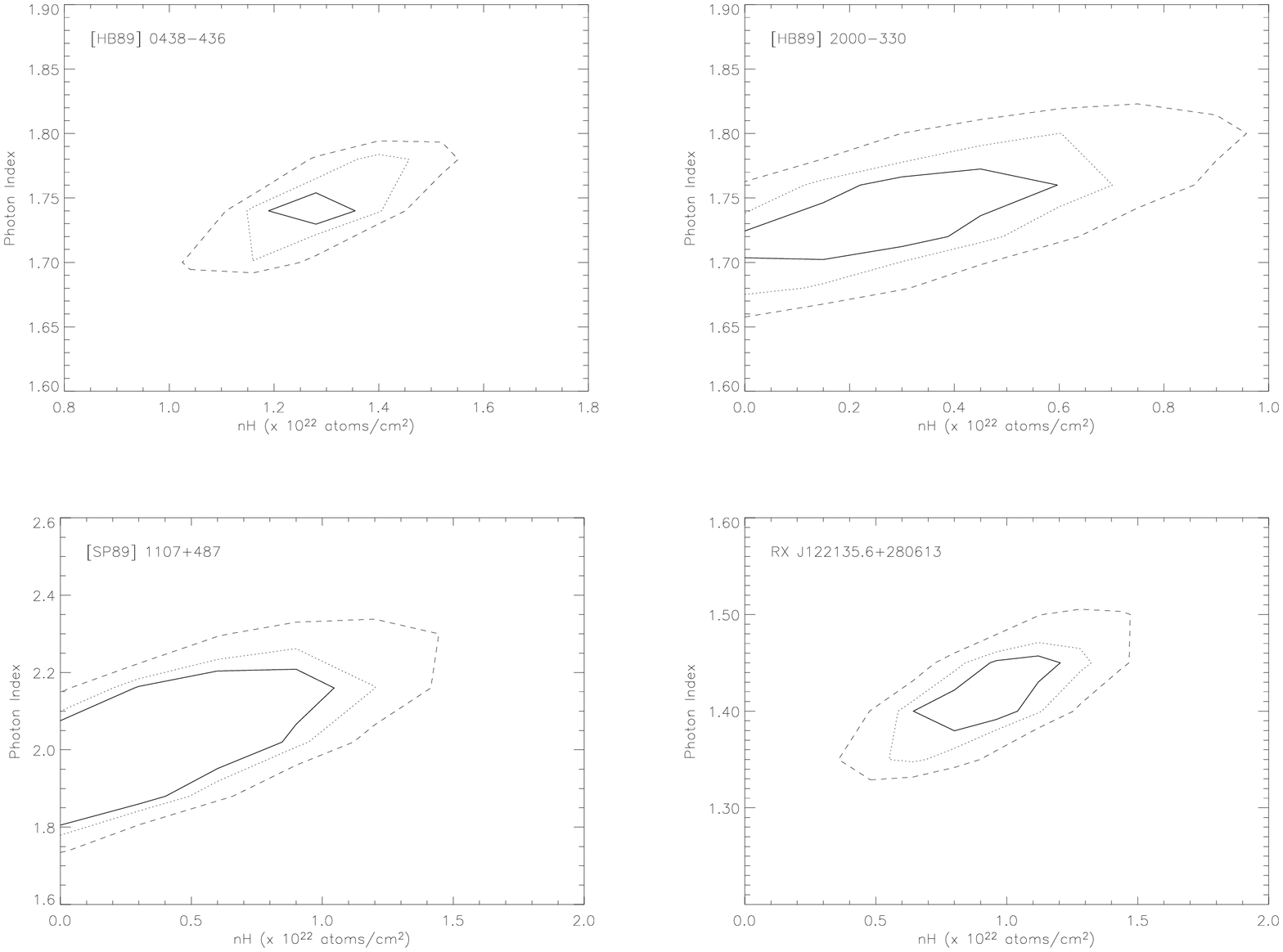,angle=0,width=16cm}
\caption{$\chi^2$ contour plots for each of the four quasars. In each case the fits have been computed for the PN, MOS1 and MOS2 instruments combined (although we have also computed the fits for each instrument seperately so that the parameter overlap can be seen -- see text for details). Contours correspond to the 68, 90 and 99 $\%$ confidence intervals for two interesting parameters.}
\label{contours}
\end{center}
\end{figure*}

Fig.~\ref{contours} shows the 68, 90 and 99\% confidence contours for the power-law photon-index and the redshifted absorption column density; each plot combines the data from all three EPIC cameras. We have also compared confidence contours for each of the three instruments individually and find them to be in good agreement. They overlapped each other at the 99\% confidence level in all cases, at the 90\% level in all cases but the [HB89] 0438$-$436 PN data and at the 68\% level for [SP89] 1107+487; contours for the two MOS instruments were in good agreement to the 68\% level for all sources.

The two radio-loud quasars, [HB89] 0438$-$436 and [HB89] 2000$-$330, were fit with power-laws with similar, relatively hard power-law indices ($\Gamma\sim1.7$ in each case). This is in agreement with previous authors who have analysed {\sl ASCA} and {\sl ROSAT} spectra (e.g. Elvis et al. 1994; Bechtold et al. 1994b Cappi et al. 1997; Reeves \& Turner 2000). In both objects there appears to be an absorption component present in addition to the Galactic value; this is required for [HB89]~0438$-$436 ($\Delta\chi^2=298.3$; see above) but not for [HB89] 2000$-$330 ($n_H \le 0.95\times 10^{22}$ with 99\% confidence; see above). Again, this is consistent with previous results.

The two radio-quiet sources, [SP89] 1107+487 and RX J122135.6+280613, have not been as well-studied previously as the two radio-loud sources and the spectrum of RX J122135.6+280613 had not been obtained until now. [SP89] 1107+487 is the softest of our four sources ($\Gamma\sim 2$) and the error range is close to that of Bechtold et al. (1994b; a redshifted source of absorption may not be required in addition to the Galactic value although it is not ruled out ($n_H \le 1.5 \times 10^{22}$ atoms/cm$^2$ with 99\% confidence; see above). RX J122135.6+280613 is the hardest of our sources ($\Gamma\sim 1.4$; see above) and does appear to require additional absorption ($\Delta\chi^2=28.3$; see above). Thus the spectra of these four quasars might suggest that the spectral hardness is dependent on the degree of intrinsic absorption taking place at lower X-ray energies; the similarity in the photon indices of the two radio-loud quasars, however, would suggest that the absorption is not the only factor contributing to spectral hardness. 

All spectra were fit very well with absorbed power-law models and did not require the presence of black-body, iron line or reflection components. Additional fits were attempted, this time including only a power-law, Galactic absorption and an iron line positioned at one of 6.4, 6.7 and 6.97 keV in the rest frame (the range of energies searched for iron lines is indicated on each residuals panel in Fig.~\ref{spectra} by a horizontal line). Fits were performed for assumed narrow (0.01 keV) and broad (0.1 keV) lines and the resultant upper limits ($99\%$ confidence based on one interesting parameter) to the equivalent width (in the rest frame of the quasar) are listed in Table~\ref{lines}. There are no previous such estimates for the two radio-quiet sources but [HB89] 0438$-$436 and [HB89] 2000$-$330 are included in the {\sl ASCA} sample of Reeves \& Turner (2000); the upper limits obtained by these authors are less well constrained but, as in the case of the {\sl XMM} data, there was no suggestion of any significant iron line emission. The upper limits, however, are sufficiently large compared with equivalent widths of lines in Seyfert galaxies (e.g. Nandra et al. 1997) that we are unable to rule out entirely the presence of any lines at Seyfert levels.

Further fits were made using the {\sc pexriv}/{\sc pexrav} models so as to determine upper limits to the flux from a possible component due to reflection from an ionized accretion disc. In all four cases the reflection component was consistent with zero but 99\%  confidence upper limits permitted values of up to 0.5, 0.8 and 0.7 for the reflection scaling factor (the ``rel-refl'' parameter in the models) for [HB89] 0438$-$436, [HB89] 2000$-$330 and RX J122135.6+280613 respectively; [SP89] 1107+487 was unconstrained. In each case the ionization parameter was also unconstrained, with error bars many times larger than the associated ionization.

\section{Broadband Spectra}

\begin{figure*}
\begin{center}
\leavevmode
\psfig{file=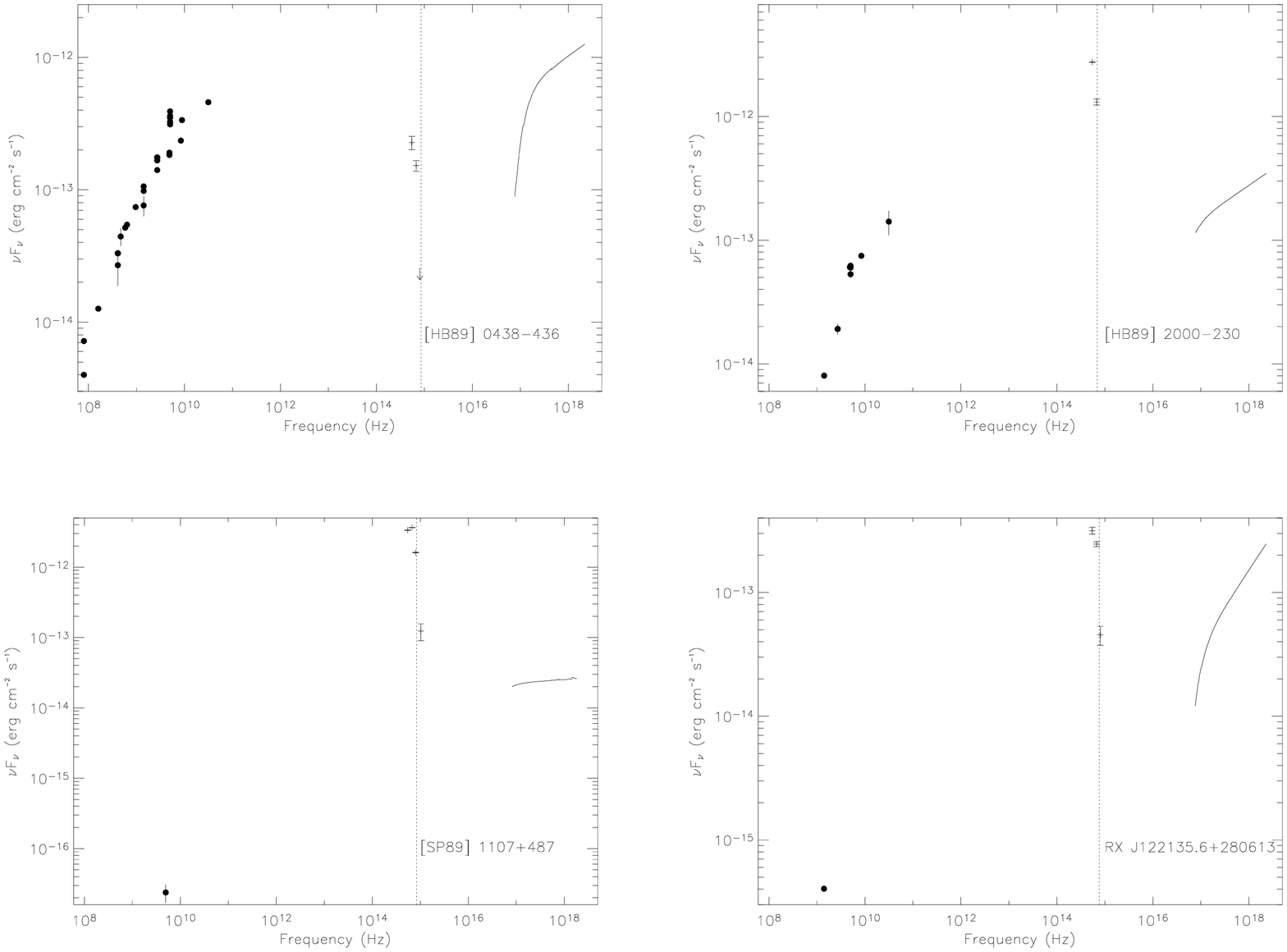,angle=0,width=14cm}
\caption{Observed frame broadband spectra plotted in $\nu F_{\nu}$--space. The optical data have been corrected for Galactic absorption using the Galactic reddening estimates of Schlegel et al. (1998) and the extinction curve of Cardelli et al. (1989). The X-ray data have been corrected using {\sc xspec}. We have also added archival radio data from the literature (obtained from NED and references therein). The vertical dotted lines correspond to the shifted 912 \AA~ Lyman break for each quasar.}
\label{nufnu}
\end{center}
\end{figure*}

Fig.~\ref{nufnu} shows the broadband spectra and model fits plotted in $\nu f_{\nu}$--space. We used the Galactic reddening estimates of Schlegel, Finkbeiner \& Davis (1998) (giving $E(B-V)=$ 0.014, 0.130, 0.019 and 0.022 magnitudes for [HB89] 0438$-$436, [HB89] 2000$-$330, [SP89] 1107+487 and RX J122135.6+280613 respectively), the extinction curve of Cardelli, Clayton, Mathis (1989) and the Galactic $n_H$ values (see Table\ref{quasars}) to correct the data for Galactic absorption. 

The two quasars which have significant redshifted absorption, [HB89] 0438$-$436 and (to a lesser extent) RX J122135.6+280613, have much greater proportions of X-ray flux (relative to the optical) compared with [HB89] 2000$-$330 and [SP89] 1107+487. The lack of absorption in these latter two sources makes it clear that the X-ray power law of [HB89] 2000$-$330 is much harder than the almost flat spectrum of [SP89] 1107+487. The flux of the optical data for each of the four sources appears to show a turnover which corresponds to the position of the Lyman break (as marked by the vertical dotted line in Fig.~\ref{nufnu}). For completeness we also include archival radio data which have been obtained from the literature (NASA/IPAC Extragalactic Database (NED) and references therein).

The similarity in gradient of the (unabsorbed) X-ray and radio curves for the two radio-loud objects is striking. In particular we note that for [HB89] 2000$-$330, the source with no significant intrinsic absorption, the $V$ band point also appears aligned with the radio data (at frequencies higher than the GHz peak). We find that the spectral index (i.e. $S_{\nu}\propto\nu^{\alpha}$ where $S_{\nu}$ is the flux density at frequency $\nu$ and $\alpha$ is the spectral index) of the radio--optical data agrees with that of the X-ray data to within $7\%$ and has value $\sim -0.7$. This would suggest some direct connection between the radio, optical and X-ray-emitting regions and we discuss this further in Section 6.

\section{Discussion}

To date, {\sl XMM-Newton} spectra for just six high ($z>2.5$) redshift quasars have been obtained and analysed in detail, in addition to the spectra of the four sources studied here. Reeves et al. (2001) observed the radio-loud source, PKS 0537$-$286 ($z=3.104$), and found it to have a very hard ($\Gamma=1.27\pm0.02$) spectrum, with weak iron and reflection emission and no significant intrinsic absorption. It was suggested that the high X-ray luminosity was due to inverse Compton emission due to the upscattering of either jet or disc photons. Ferrero \& Brinkmann (2003) observed two radio-loud (PKS~2126$-$158, PKS~2149$-$306; $z=3.27 \mbox{ and } 2.34$ respectively) and two radio-quiet (Q~0000$-$263, Q~1442+2931  $z=2.64 \mbox{ and } 4.10$ respectively) quasars. The radio-loud sources had very flat spectra ($\Gamma\le 1.5$) and PKS~2126$-$158 required a redshifted absorption component ($1.40\times 10^{22}\mbox{ cm}^{-2}$) in addition to the Galactic value. The two radio-quiet sources had steeper spectra ($\Gamma\sim2$) and neither required additional absorption. Finally Grupe et al. (2004) presented {\sl XMM-Newton} observations of a radio-loud object, RX~J1028$-$0844 ($z=4.276$), and a radio-quiet object, BR~0351$-$1034 ($z=4.351$); their spectra were fit with power-laws with photon indices $\Gamma\sim1.3$ and $\sim2$ respectively. The presence of intrinsic absorption in RX~J1028$-$0844 (as suggested by previous {\sl ASCA} observations) was not confirmed and the observations did not support the proposed correlation between spectral parameters and redshift. Similarly the observation of BR 0351$-$1034 did not support previous claims suggesting that high redshift radio-quiet quasars have weaker X-ray emission.  

Our {\sl XMM-Newton} observations of high redshift quasars are consistent with the general trends found from observations with previous telescopes but the fit parameters have been determined to a much greater accuracy. Both radio-loud sources have relatively hard spectra ($\Gamma\sim 1.7$) and significant intrinsic absorption is required to fit the spectrum of [HB89] 0438$-$436; we note that this is just the second radio-loud quasar observed by {\sl XMM-Newton} to show intrinsic absorption (see discussion in Grupe et al. 2004). The X-ray spectra of the radio-quiet objects are either soft and unabsorbed (as in the case of [SP89] 1107+487 and the two radio-quiet sources observed by Ferrero \& Brinkmann 2003) or hard and absorbed (as for RX J122135.6+280613).

\begin{table}
\caption{Results of additional fits including an iron line. The presence of either broad or narrow lines was searched for at each of three positions: 6.4, 6.7 and 6.97 keV (in the rest frame of the quasar; the equivalent observed frame energies are given in paretheses). Upper limits (99 \% confidence) to the rest frame equivalent width are listed.}
\begin{tabular}{lccc}
\hline
\hline
Source&Line&Width&EW\\
&(keV)&(keV)&(eV)\\
\hline
\mbox{[HB89]} 0438$-$436&6.40 (1.66)&0.01&$<120$\\
&6.40 (1.66)&0.1 &$<250$\\
&6.70 (1.74)&0.01&$<120$\\
&6.70 (1.74)&0.1 &$<250$\\
&6.97 (1.81)&0.01&$<140$\\
&6.97 (1.81)&0.1 &$<270$\\
\hline
\mbox{[HB89]} 2000$-$330&6.40 (1.34)&0.01&$<73$\\
&6.40 (1.34)&0.1 &$<170$\\
&6.70 (1.40)&0.01&$<73$\\
&6.70 (1.40)&0.1 &$<200$\\
&6.97 (1.46)&0.01&$<142$\\
&6.97 (1.46)&0.1 &$<230$\\
\hline
\mbox{[SP89]} 1107+487&6.40 (1.62)&0.01&$<480$\\
&6.40 (1.62)&0.1 &$<860$\\
&6.70 (1.69)&0.01&$<680$\\
&6.70 (1.69)&0.1 &$<800$\\
&6.97 (1.76)&0.01&$<650 $\\
&6.97 (1.76)&0.1 &$<760$\\
\hline
RX J122135.6+280613&6.40 (1.49)&0.01&$<220$\\   
&6.40 (1.49)&0.1 &$<460$\\
&6.70 (1.56)&0.01&$<300$\\
&6.70 (1.56)&0.1 &$<480$\\
&6.97 (1.62)&0.01&$<240$\\
&6.97 (1.62)&0.1 &$<480$\\
\hline
\label{lines}
\end{tabular}
\end{table}

We also compare the values obtained for photon index and redshifted absorption with those of samples chosen by other authors. The photon index values for the two radio-loud objects are in agreement with previous samples ($\Gamma=1.61\pm0.04$; Cappi et al. 1997 and references therein). The two radio-quiet sources, however, have photon indices falling either side of the range predicted ($\Gamma=1.67\pm0.11$; Vignali et al. 1999 and references therein), although this may be related to the absorption as discussed above. In a sample of 62 quasars observed by {\sl ASCA}, 11 of the 21 sources at $z>2$ were modeled with a redshifted absorption component of $10^{22}$--$10^{23}$ cm$^{-2}$; two of these objects were radio-quiet (Reeves \& Turner 2000). Thus the two sources in our sample showing redshifted absorption appear to comply with the proposed correlation between column density and redshift; the upper limits to the absorption for the other two sources also fall in a similar range and likewise the values obtained by Ferrero \& Brinkmann (2003) for their two radio-loud objects. Conversely the estimates of Cappi et al. (1997) were lower, as discussed by these authors and also by Reeves \& Turner (2000). 

Given the high redshift of these objects and the energy range of the observations, it is perhaps to be expected that we would not detect the presence of any black-body component, relating to the accretion disc that is typically assumed to be present around the central black hole of a quasar (e.g. Gondhalekar, Rouillon-Foley, Kellet 1996). We have estimated the rest frame luminosity (for a WMAP cosmology) of each source in the 2--30 keV range and find that the values ($\sim 0.5\times 10^{46}$ -- $1.6\times 10^{47}$ erg s$^{-1}$; see Table~\ref{model}) are consistent with high mass black holes ($\ge$ 0.04 -- 1.2 $\times 10^9 M_{\odot}$), assuming that the Eddington limit is valid. For the unabsorbed source [SP89] 1107+487 we can also obtain an upper limit to the temperature of an assumed blackbody component ($\le40$ eV) which corresponds to a lower limit on the black hole mass of $\ge 0.02 \times 10^9 M_{\odot}$. We use the mass estimate of each source to derive the corresponding inner disc temperature ($\le$ 16 -- 38 $\times 10^4$ K; following King \& Puchnarewicz 2002) and find that it corresponds to optical/ultraviolet emission. It is therefore not surprising that the X-ray spectra are relatively hard compared with Seyfert-like objects; there is no X-ray contribution to the disc emission due to the apparently truncated inner disc. Conversely the softer spectrum of [SP89] 1107+487 is more comparable with that of a Seyfert galaxy although still shows no evidence for an accretion disc component in the X-ray emission.

The presence of narrow and broad iron K$\alpha$ lines has been well-documented in Seyfert galaxies (e.g. Nandra \& Pounds 1994; Nandra et al. 1997) and is thought to arise via fluorescence in the near-neutral accretion disc surrounding the black hole (e.g. Fabian et al. 2000). Considering that it is now widely accepted that all active galaxies and quasars contain an accreting black hole it might seem surprising that we see no evidence for a narrow or broad iron line in any of our four quasars. This was also the case for the four quasars observed by Ferrero \& Brinkmann (2003) and for many of the quasars included in the {\sl ASCA} sample studied by Reeves \& Turner (2000). Similarly only two (possibly three) of the low-redshift ($z=0.06--1.96$) radio-loud objects observed by {\sl Chandra} showed significant iron lines (at rest frame energies of 6.4 keV; Gambill et al. 2003).

The absence of an iron line in some Seyfert galaxies has been typically explained in terms of orientation of the object; the orientations derived from iron line emission have not always been consistent with those derived from e.g. radio jet or broad line regions but such problems may be resolvable by considering electron-scattering from a disc atmosphere or warping of the disc (Fabian et al. 2000. and references therein). In the case of the more luminous quasars the absence of an iron line and reflection features may further be explained in terms of a much higher accretion rate and subsequently higher ionization (e.g. Fabian et al. 2000) and there is some tentative evidence from {\sl Ginga} and {\sl ASCA} data for an X-ray ``Baldwin'' effect, with the EW of the iron line decreasing with increasing luminosity (Iwasawa \& Taniguchi 1993; Nandra 1999). However, we might expect to see a greater range of ionization states, showing some degree of correlation with accretion rate if this were to be the case; confirmation of the effect via larger samples of higher resolution observations is required.

Similarly the absence of a Comptonized reflection bump provides further problems for this standard model of accretion onto a black hole. With a predicted rest-frame energy of $\sim 30$ keV, the peak of the reflection bump (or at least some significant curvature) should be observable in {\sl XMM-Newton} spectra for these high redshift objects. A single power-law should not be able to provide such a good fit as seen in the four quasars studied here (and similarly in e.g. an {\sl ASCA} sample of radio quiet quasars with $z=2$ studied by Vignali et al. 1999). In order to test this further we have attempted to fit a broken power-law to each dataset; if a reflection bump were present then we would expect to see some form of ``break'' at or close to a rest frame energy of $\sim 30$ keV. This was not the case for any of our quasars (and we confirm that the high energy part of each {\sl XMM-Newton} spectrum was source-dominated with no evidence for a problem with the background subtraction). We have also attempted to fit the {\sc pexriv} model and obtained only upper limits to the reflection scaling factor (i.e. ratio of reflected to direct components). 

The absence of any ``bump'' in the spectrum of these high luminosity quasars might suggest that the process by which a bump is produced (e.g. reflection?) in low luminosity objects is not yet well-understood. Potential explanations include the relatively high power-law contribution, a possible low covering factor of the reflector (as seen by the emitting source) and the fact that the critical energy for the bump falls in the less sensitive region of the detector (i.e. 5--10 keV in the observed frame) where there are fewer counts. If we were to observe up to 100 keV (in the rest frames) then the higher-energy flattening and fall-off of any bump due to reflection should also be visible. Alternatively the large inner disc radius (and thus a potentially lower albedo) thought to be present in these high luminosity systems may result in a less significant reflection bump than expected.


While we are unable to rule out the presence of any iron line or reflection component at Seyfert levels, there are other possible models which could produce similar power-law spectra. In particular we note that all four of our sources are radio-emitting and therefore that a jet is likely to be present. The jet nature of [HB89] 0438$-$436 has been imaged (e.g. Shen et al. 1998; Tingay et al. 2002) and since [HB89] 2000$-$330 is a gigahertz peaked-spectrum source (O'Dea, Baum \& Stanghellini 1991), it is likely to display a compact radio jet, requiring Very Long Baseline Interferometry (VLBI) in order to resolve it (e.g. Stanghellini et al. 1997). Jet-like natures for radio-quiet quasars are both predicted (e.g. Falcke \& Biermann 1995) and observed (e.g. Blundell \& Beasley 1998), although the low flux density of [SP89] 1107+487 would make it particularly difficult to resolve with current telescopes. However, the presence of jets can also be inferred from the radio spectrum (e.g. Blandford \& K\"onigl 1979) and this method has been used to predict a jet in the X-ray binary source Cygnus X-1 (Fender et al. 2000) which was confirmed independently using VLBI (Stirling et al. 2001).

Jets are well-known for emitting at radio wavelengths but their effects at other frequencies are often not taken into account, particularly during analysis of X-ray data. Their emission is produced via the synchrotron mechanism which requires a population of relativistic electrons, as does the Comptonizing corona (which is thought to produce the power-law emission). The observed correlations between the X-ray power-law emission and radio jet of black hole X-ray binaries in the low/hard state suggests a close relationship between these two populations of relativistic electrons (e.g. Corbel et al. 2003 and references therein). The jets are also sufficiently powerful that they can emit at higher frequencies, with some being resolved by {\sl Chandra} (Siemiginowska et al. 2002; Corbel et al. 2002; see also Harris \& Krawczynski 2002). For example, study of the multiwavelength variability and broadband spectrum has suggested that the synchrotron radio spectrum of the jet-emitting Seyfert 1 galaxy III Zw 2 extends into the optical/UV/X-ray regime (Salvi et al. 2002). Similarly Zamorani et al. (1981) found that the X-ray luminosity of radio-loud AGN is a factor of $\sim 3$ larger than that of radio-quiet objects, suggesting some sort of direct correlation between the X-ray and radio emitting regions. The similarities in spectral index between the radio-optical and X-ray  datasets might suggest a common jet origin for the high energy photons in [HB89] 2000$-$330 (and the other sources?) also; the spectral index $\sim -0.7$ is consistent with optically thin synchrotron emission.

The mechanism by which the jet X-rays are emitted is not yet well-understood but it is almost certainly non-thermal. Falcke \& Biermann (1995) modelled the broadband spectrum with synchrotron self-Comptonization of jet photons producing a significant X-ray contribution. Alternatively Markoff et al. (2001, 2003) were able to fit a model to the {\em hard} state of the X-ray binaries XTE J1118+408 and GX~339$-$4, in which synchrotron emission dominated at X-ray energies; the model was able to predict the observed X-ray/radio correlations as well as the spectrum. A final source of non-thermal X-ray emission is the inverse-Compton scattering by {\em jet} electrons of external sources of emission, such as the accretion disc, surrounding gas and dust and the cosmic microwave background as considered by e.g. Dermer \& Schlikeiser (2002 and references therein) and Siemiginowska et al. (2002). All these mechanisms produce a power-law spectrum which can potentially dominate over any reflection component at the energies observed in our {\sl XMM-Newton} observations. Concentrated multiwavelength observing campaigns, so as to obtain broadband spectra in the context of these models, would be extremely valuable.

\section{Conclusions}
We have obtained {\sl XMM-Newton} spectra of two radio-loud ([HB89] 0438$-$436, [HB89] 2000$-$330) and two radio-quiet ([SP89] 1107+487, RX J122135.6+280613) quasars. All spectra could be well-fit by simple absorbed power-laws, with [HB89] 0438$-$436 and RX J122135.6+280613 requiring redshifted absorption contributions in addition to the Galactic component. The spectra did not require the addition of iron line or reflection components and we suggest that these lower-temperature discs may have a significantly lower albedo; alternatively the derived upper limits still allow these features to be present at Seyfert-like (low luminosity) levels. However, we remind readers that inverse-Compton scattering of photons off electrons in a corona is not the only feasible mechanism by which a power-law at X-ray energies may be produced. Since all four of our sources have been detected in the radio and are therefore possible jet sources, we suggest that jet models are considered as feasible alternatives.

\section*{acknowledgements}
This research has made use of the NASA/IPAC Extragalactic Database (NED) which is operated by the Jet Propulsion Laboratory, California Institute of Technology, under contract with the National Aeronautics and Space Administration. We thank the anonymous referee for the very detailed and helpful report.

\end{document}